\documentclass[twocolumn,showpacs,preprintnumbers,amsmath,amssymb,superscriptaddress]{revtex4}
\usepackage{graphicx}
\usepackage{dcolumn}
\usepackage{bm}

\begin{document}

\preprint{}

\title{Consequences of imperfect mixing in the Gray-Scott model}

\author{M.-P. Zorzano }
\email{zorzanomm@inta.es} \homepage{http://www.cab.inta.es {\it Present
    address: CAB}}

\affiliation{Centro de Astrobiolog\'{\i}a (CSIC-INTA), Carretera de
Ajalvir km 4,    28850 Torrej\'{o}n de Ardoz, Madrid, Spain}
\author{D. Hochberg}%
\email{hochberg@laeff.esa.es} \affiliation{Centro de
Astrobiolog\'{\i}a (CSIC-INTA), Carretera de Ajalvir km 4,    28850
Torrej\'{o}n de Ardoz, Madrid, Spain}
\author{F. Mor\'{a}n}
\email{fmoran@solea.quim.ucm.es} \affiliation{Centro de
Astrobiolog\'{\i}a (CSIC-INTA), Carretera de Ajalvir km 4,    28850
Torrej\'{o}n de Ardoz, Madrid, Spain} \affiliation{ Departamento de
Bioqu\'{\i}mica y Biolog\'{\i}a Molecular, Facultad de Ciencias
Qu\'{\i}micas, Universidad Complutense de Madrid, Spain
}%

\date{\today}

\begin{abstract}
We study  an autocatalytic reaction-diffusion scheme, the Gray-Scott
model, when the mixing processes do not homogenize the reactants.
Starting from the master equation, we derive the resulting coupled,
nonlinear, stochastic partial differential equations that {\em
rigorously} include the spatio-temporal fluctuations resulting from
the interplay between the reaction and mixing processes. The fields
are complex and depend on correlated complex noise terms. We
implement a novel method to solve for these {\em complex} fields
numerically and extract accurate information about the system
evolution and stationary states under {\em different mixing
regimes}. Through this example, we show how the reaction induced
fluctuations interact with the temporal nonlinearities leading to
results that differ significantly from the mean-field (perfectly
mixed) approach. This procedure can be applied to an arbitrary
non-linear reaction diffusion scheme.

\end{abstract}


\maketitle

Reaction diffusion networks are relevant to a broad scope of
chemical, biological and nonlinear optical systems where the agents
must diffuse about before they meet and react \cite{RMP}. To date,
much of the theoretical work devoted to the analysis of reaction
network dynamics has been based on the assumption that the system is
``well mixed'', {\it i.e.} that the concentration of each species is
always uniform throughout the system.  This regime is then solved
using mean-field approximations which may be adequate under perfect
stirring conditions or equivalently when the diffusion rates are
very high. But if the reaction rates are nonlinear, which is the
case of the ubiquitous autocatalytic and catalytic networks, any
departure from the assumption of perfect mixing can have major
dynamical consequences. There are numerous experimental evidences on
the outcome of catalytic processes under imperfect mixing
\cite{Nature}. For instance, it has been experimentally shown that
imperfect mixing in autocatalytic systems can lead to chemical
reactions that occur at seemingly random intervals \cite{clocks},
crystallization processes that yield sometimes all left-handed and
sometimes all right-handed crystals \cite{kondepudi} and reactions
in which changing the rate at which a solution is stirred can cause
a transition from a stationary, time-independent state to one of
periodic or even chaotic oscillation \cite{Chaos}. These phenomena
have been studied primarily in inorganic chemical media, but there
are implications also for relevant biological systems: hypercycle
networks, viral quasispecies dynamics, and ecosystems, etc.
\cite{RDSpatEc,LNature,TPB}. There is a wealth of theoretical works
related to the effects of fluctuations on these systems that
intuitively show the relevance that density fluctuations may have on
stability, pattern formation and the effects of dimensionality
\cite{GO}. However, in these approaches, the fluctuations are simply
added {\it ad-hoc} to the mean-field description and the noise
strength and correlations are chosen arbitrarily.

In this Brief Report we address this issue and work through a
general-purpose, mathematical and numerical procedure to study the
mixing process in reaction-diffusion problems. The mixing process in
nonlinear reaction systems has been described mostly for simple
reaction schemes \cite{TW} but has never been fully exploited to
solve for the Langevin-type equation, since the fluctuations in the
general case, turn out to be complex. Our approach is based on (1)
the application of a standard, field-theoretic method to
characterize the reaction induced fluctuations and derive a
Langevin-type description of the underlying molecular dynamics, (2)
the non-dimensionalization of the problem to characterize the
dependence on the agents' diffusivities, reaction rates and the
spatial dimension and (3) the separation of the real and imaginary
parts of the noise to solve numerically for the real and imaginary
parts of the stochastic fields. As an example of this consistent
description, we treat the Gray-Scott (GS) model in two spatial
dimensions \cite{Gray83} and {\em vary the diffusion rates to
rigorously explore different mixing regimes}. We choose this model
to facilitate visualizing the differences between perfect and
imperfect mixing as the GS model is known to exhibit spatial pattern
formation and to be very sensitive to multiplicative noise
\cite{us}. This procedure can be applied to an arbitrary reaction
scheme for any spatial dimension. The method presented here allows
one to solve for the {\em full set of complex Langevin equations}
and to extract information about the spatio-temporal time evolution
of the system and its stationary states under different mixing
regimes. We show numerically that only the noise averaged real parts
of the complex fields correspond to the density of the reactants. We
compare our results with the mean-field case, which is recovered
only in the infinite diffusion limit.

The GS model, corresponds to the following chemical reactions:
\begin{eqnarray}\label{reaction}
U + 2 V &\stackrel{\lambda}{\to}& 3 V, \nonumber\\
V &\stackrel{\mu}{\to}& p, \nonumber\\
U &\stackrel{\nu}{\to}& q, \nonumber\\
&\stackrel{u_0}{\to}& U.
\end{eqnarray}
The concentrations of the chemical species $V$ and $U$ are functions
of $d$-dimensional space $\vec{x}$ and time $t$. $\lambda$ is the
reaction rate, $p$ and $q$ are inert products, $\mu$ is the decay
rate of $V$ and $\nu$ is the decay rate of $U$. The equilibrium
concentration of $b$ is $u_0/\nu$, where $u_0$ is the feed rate
constant. The chemical species $U$ and $V$ can diffuse with
independent diffusion constants $D_u$ and $D_v$. All the model
parameters are positive.

The master equation associated with Eq.(\ref{reaction}) can be
mapped to a second-quantized description following a procedure
developed by Doi \cite{Doi}. Briefly, we introduce annihilation and
creation operators $a_i$ and $a^\dag_i$ for $V$ and $b_i$ and $b^\dag_i$ for
$U$ at each lattice site $i$, with the commutation relations
$[a_i,a_j^{\dag}] = \delta_{ij}$ and $[b_i,b_j^{\dag}] =
\delta_{ij}$. The vacuum state satisfies $a_i|0\rangle =
b_i|0\rangle = 0$. We then define the time-dependent state vector
$|\Psi(t)\rangle = \sum_{\{m\},\{n\}}P(\{m\},\{n\},t) \prod_i(
{a}_i^\dag)^{m_i}( {b}_i^\dag)^{n_i}|0\rangle$. $P(\{m\},\{n\},t)$
is the probability to find $m_i$ $U$ and $n_i$ $V$ particles at site
$i$ at time $t$. The master equation can be written as a
Schr\"{o}dinger-like equation $-\frac{\partial
|\Psi(t)\rangle}{\partial t} = {H}|\Psi(t)\rangle,$ where the
lattice hamiltonian or time-evolution operator is a function of
$a_i,a^\dag_i,b_i,b^\dag_i$ and is given by
\begin{eqnarray}\label{hamiltonian}
{H} &=& \frac{D_v}{l^2}\sum_{(i,j)}( {a_i}^\dag - {a_j}^\dag)( {a_i}
-  {a_j} ) + u_0\sum_i(1 -  {b_i}^\dag)\nonumber \\ &+&
\frac{D_u}{l^2}\sum_{(i,j)}( {b_i}^\dag -
 {b_j}^\dag)( {b_i} -  {b_j} )\nonumber \\
&-&\frac{\lambda}{2}\sum_i[( {a_i}^\dag)^3  {a_i}^2 {b_i}
- ( {a_i}^\dag)^2  {a_i}^2  {b_i}^\dag  {b_i}]\nonumber \\
&+& \nu \sum_i( {b_i}^\dag -1) {b_i} + \mu \sum_i( {a_i}^\dag -1)
{a_i} .
\end{eqnarray}
This has the formal solution $|\Psi(t)\rangle = \exp(-
{H}t)|\Psi(0)\rangle$.

The operator Eq.(\ref{hamiltonian}) is next mapped onto a continuum
field theory. This procedure is now standard and we refer to
\cite{Peliti} for further details . For the GS system, we obtain the
path integral $\exp(- {H}t) = \int \mathcal{D}a \mathcal{D}\bar{a}
\mathcal{D}b \mathcal{D}\bar{b}\, e^{-S[a,\bar{a},b,\bar{b}]}$, over
the continuous (and generally complex) stochastic fields
$a(\vec{x},t),\bar a(\vec{x},t), b(\vec{x},t), \bar b(\vec{x},t)$
where the action $S$ is given by
\begin{eqnarray}\label{action}
S &=& \int d^dx \int_0^{\tau} dt [\bar{a}\partial_t a + D_v\nabla
\bar{a} \nabla a + \bar{b}\partial_t b + D_u\nabla \bar{b} \nabla
b\nonumber \\
&+& \mu(\bar{a} - 1)a + \nu(\bar{b} - 1)b -u_0(\bar{b} - 1)\nonumber
\\ &-&\frac{\lambda}{2}(\bar{a}^3a^2b - \bar{a}^2a^2\bar{b}b) ].
\end{eqnarray}
We have omitted terms related to the initial state. Apart from
taking the continuum limit, the derivation of this action is exact,
and in particular, no assumptions regarding the precise form of the
noise are required. For the final step, we perform the shift
$\bar{a} = 1 + a^*$ and $\bar{b} = 1 + b^*$ on the action $S$. We
represent the terms quadratic in $a^*, b^*$ by an integration over
Gaussian noise terms, which allows us to then integrate out the
conjugate fields \footnote{This can be done provided we ignore the
cubic terms which would give information on higher order cumulants
of the fluctuations.}. To proceed, we note that $e^{(\lambda
a^2b({a^*}^2 - a^*b^*))} \approx \int \mathcal{D}\xi \mathcal{D}\eta
\, P(\xi,\eta)\,e^{(a^*\xi + b^* \eta)}$, where the noise functions
$\xi,\eta$ are distributed according to a double Gaussian as
$P(\xi,\eta) = \exp\big(-(\xi,\eta)V^{-1} \left(\begin{array}{c}
  \xi \\
   \eta \\
\end{array} \right)
\big)$, with $V$ the matrix of noise-noise
correlation functions
\begin{equation}\label{matrix}
\qquad V =
\left(%
\begin{array}{cc}
  \langle \xi \xi \rangle & \langle \eta \xi \rangle \\
  \langle \xi \eta \rangle & \langle \eta \eta \rangle \\
\end{array}%
\right).
\end{equation}
Integrating out the conjugate fields $a^*$ and $b^*$ from the
functional integral for this shifted action then leads to the pair
of coupled nonlinear Langevin equations. We define the dimensionless
fields, $u=\frac{\lambda}{\mu}b$ and $v=\frac{\lambda}{\mu}a$,
dimensionless time, $\tau=\mu t$, and spatial coordinates
$x_{i}=\sqrt{\mu/D_u}\ \hat{x}_{i}$, $i=1,..,d$. The equations
describing the field dynamics read:
\begin{eqnarray}\label{GSN}
\partial_\tau v(\hat{x}_i,\tau) &=& \frac{D_v}{D_u}
\nabla^2 v - v + v^2u + \xi(\hat{x}_i,\tau) \\\nonumber
\partial_\tau u(\hat{x}_i,\tau) &=& \nabla^2 u -\frac{\nu}{\mu} u -
v^2u + \frac{u_0}{\mu} + \eta(\hat{x}_i,\tau),
\end{eqnarray}
with noise correlations
\begin{eqnarray}\label{noisecorr}
\langle\xi(\hat{x}_i,\tau)\rangle &=& \langle \eta(\hat{x}_i,\tau)
\rangle = 0 \nonumber
\\
\langle \xi(\hat{x}_i,\tau)\xi(\hat{x}_i',\tau')\rangle
&=& \epsilon\ v^2u \ \delta^2(\hat{x}_i-\hat{x}_i')\delta(\tau-\tau') \nonumber \\
\langle \xi(\hat{x}_i,\tau)\eta(\hat{x}_i', \tau')\rangle
&=& -\frac{\epsilon}{2} \ v^2u\ \delta(\hat{x}_i-\hat{x}_i')\delta(\tau-\tau') \nonumber \\
\langle \eta(\hat{x}_i,\tau)\eta(\hat{x}_i',\tau')\rangle &=& 0.
\end{eqnarray}
and noise strength
$\epsilon=\sqrt{\lambda}\frac{{\mu}^{(d-1)/2}}{{D_u}^{d/2}}$. In
particular, for $d=2$ dimensions, $\epsilon=\frac{\sqrt{\lambda
\mu}}{D_u}$.  The non-dimensionalization of the problem allows us to
characterize explicitly the dependence on the agents diffusivities,
reaction rates and spatial dimension. For a fixed reaction rate, as
we will see below, the transition from perfect to the imperfect
mixing regime is associated with an increase in $\epsilon$. Notice
that $\eta$ has zero autocorrelation but non-zero cross-correlation
with $\xi$, indicating that this noise is {\em complex}. Using the
Cholesky decomposition, we can express these complex noise
components as a linear combination of two uncorrelated (real) white
Gaussian noises $\theta_{1}$, $\theta_{2}$ thus:
\begin{eqnarray}\label{noise1}
\xi(\hat{x}_i;\tau)&=& v \sqrt{\epsilon u}\ \theta_{1}(\hat{x}_i;\tau)\\\label{noise2}
\eta(\hat{x}_i;\tau)&=& -\frac{1}{2} v \sqrt{\epsilon u}\ \theta_{1}(\hat{x}_i;\tau)+i\
\frac{1}{2}v  \sqrt{\epsilon u}\ \theta_{2}(\hat{x}_i;\tau).
\end{eqnarray}
Thus $u$ and $v$ are also complex fields. Through this procedure we
can separate the real and imaginary parts of the noise and solve
numerically the stochastic non-linear reaction diffusion equations
(\ref{GSN}) for the real and imaginary parts of the fields,
\cite{CPL}.  Now we can obtain numerical information from these
complex densities. We expect the imaginary parts of these fields to
be zero on average, since the stochastic averages $\langle u
\rangle$ and $\langle v\rangle$ correspond to the physical reactant
densities, \cite{HT}.

In the mean-field approximation, this system possesses three
homogeneous solutions: one absorbing state
$R=(u=\frac{u_o}{\nu}\frac{\lambda}{\mu}, v=0)$ and two non-trivial
states $B_{\pm}=(u=\frac{u_o \pm \sqrt{u_o^2 -
4\nu\mu^{2}/\lambda}}{2\nu}\frac{\lambda}{\mu},
v=\frac{1}{u}\frac{\lambda}{\mu})$. In our simulations we consider
the $d=2$ case with $\lambda=1$, $D_v/D_u=0.5$, $u_o=\nu$ and thus
the homogeneous solutions will be $R=(u=\frac{1}{\mu}, v=0)$ and
$B_{\pm}=(u=\frac{1 \pm \sqrt{1 - 4\mu^{2}/\nu}}{2}\frac{1}{\mu},
v=\frac{1}{u}\frac{1}{\mu})$. The state $B_{+}$ is globally
unstable. The trivial state $R$ is linearly stable and globally
attracting for all $\nu>0$ and $\mu>\nu$. The state $B_{-}$ is
stable if $4\mu^2 > \nu$. In a narrow region of parameter space
$(\mu,\nu)$ close to $4\mu^2=\nu$, the trivial absorbing state $R$
loses stability through a Hopf bifurcation. In the
\textit{noise-free} case, it is in the vicinity of this region where
one can find a great variety of spatiotemporal patterns in response
to a localized initial spatial perturbation \cite{Pearson}, although
no patterns are found if the initial condition is homogeneous. Here
we will consider the evolution of the system described by  Eqs.
(\ref{GSN}) with initial {\em homogeneous condition}  and subject to
the internal noise induced by the reaction, Eqs. (\ref{noise1}) and
(\ref{noise2}), in different mixing regimes.

\begin{figure}[h]
\begin{center}
\begin{tabular}{cccc}
\includegraphics[width=0.1\textwidth]{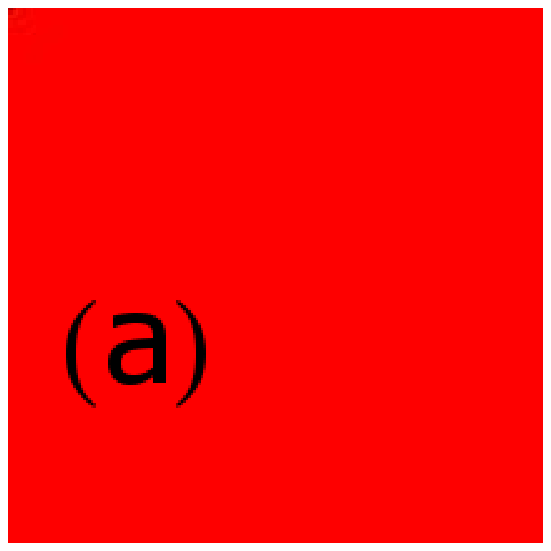} &
\includegraphics[width=0.1 \textwidth]{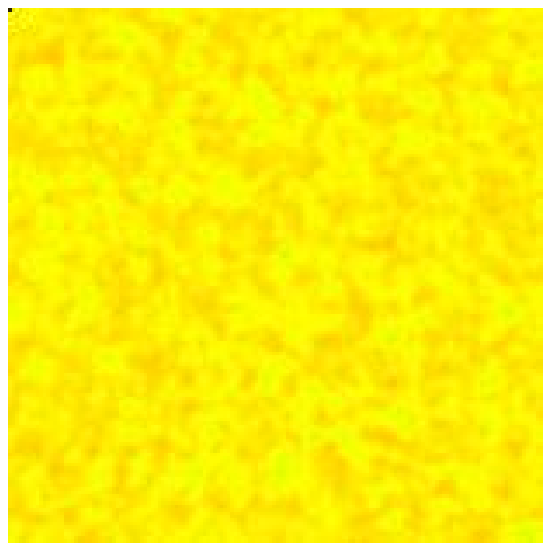} &
\includegraphics[width=0.1\textwidth]{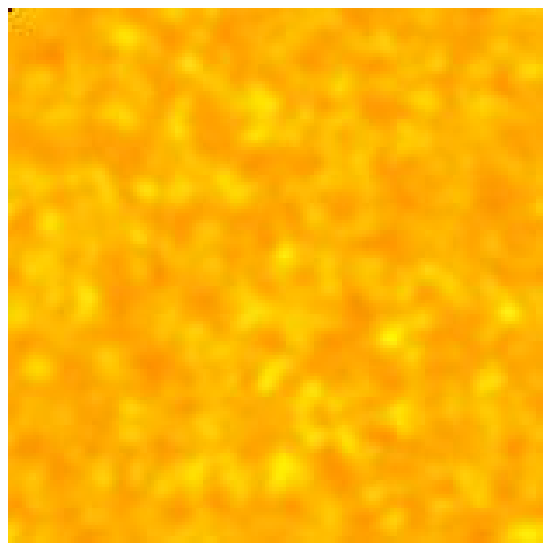} &
\includegraphics[width=0.1\textwidth]{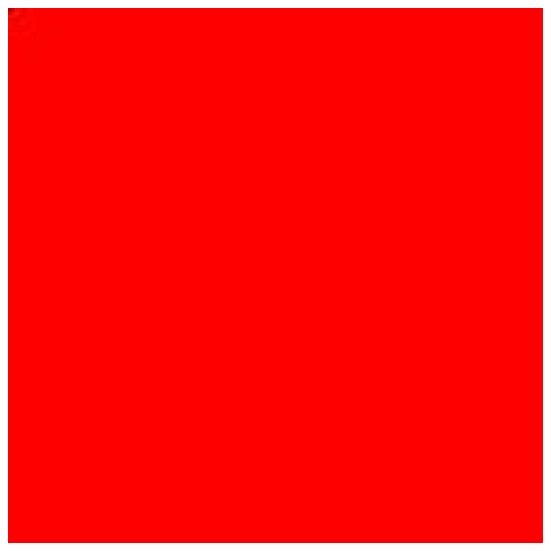}\\
 \includegraphics[width=0.1 \textwidth]{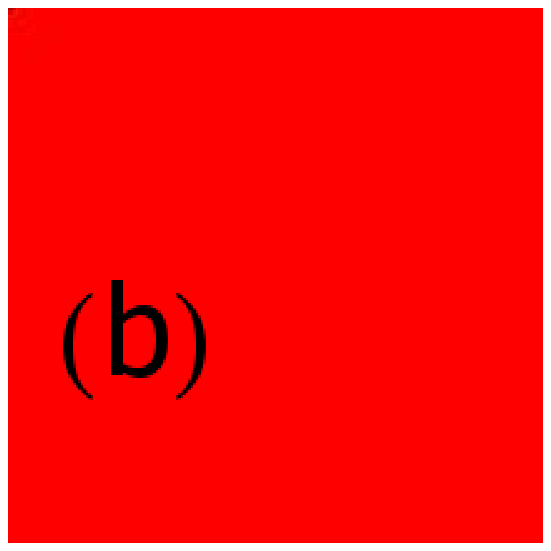}&
\includegraphics[width=0.1 \textwidth]{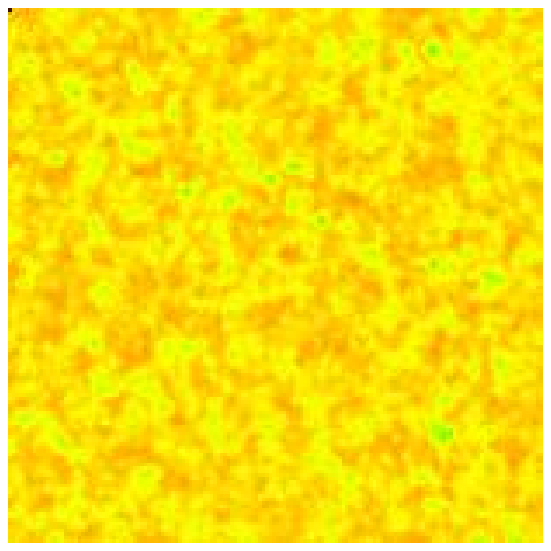} &
\includegraphics[width=0.1\textwidth]{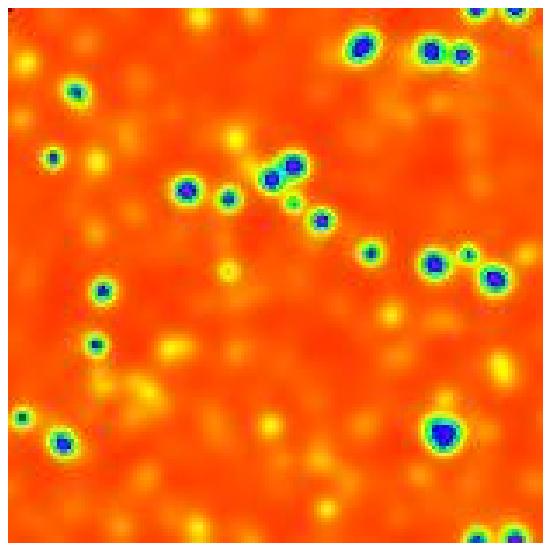} &
\includegraphics[width=0.1\textwidth]{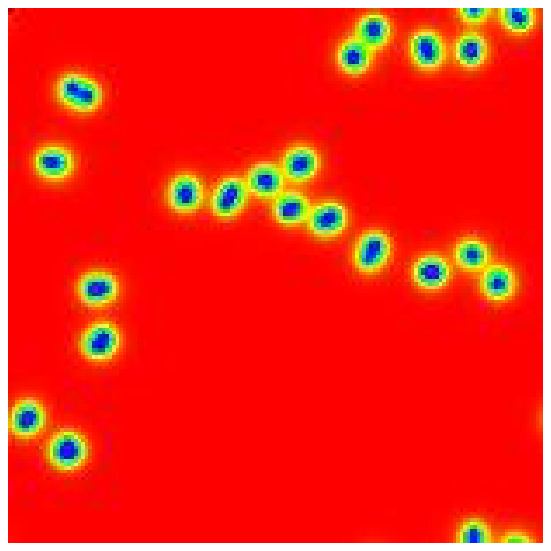}\\
\includegraphics[width=0.1\textwidth]{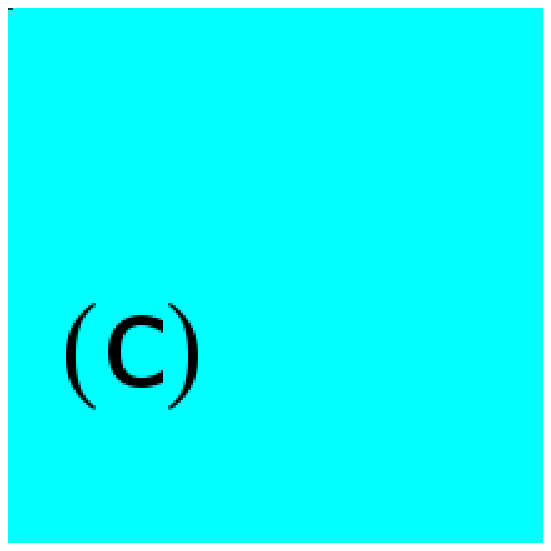}&
\includegraphics[width=0.1\textwidth]{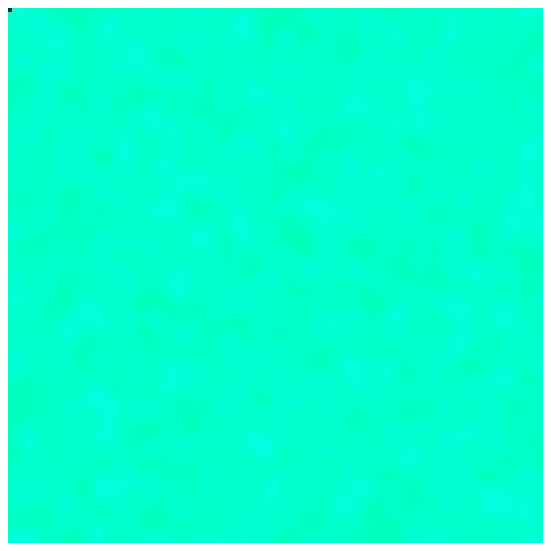} &
\includegraphics[width=0.1\textwidth]{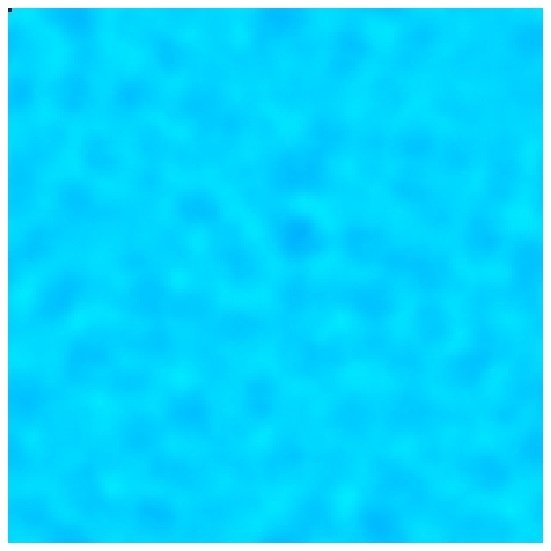} &
\includegraphics[width=0.1\textwidth]{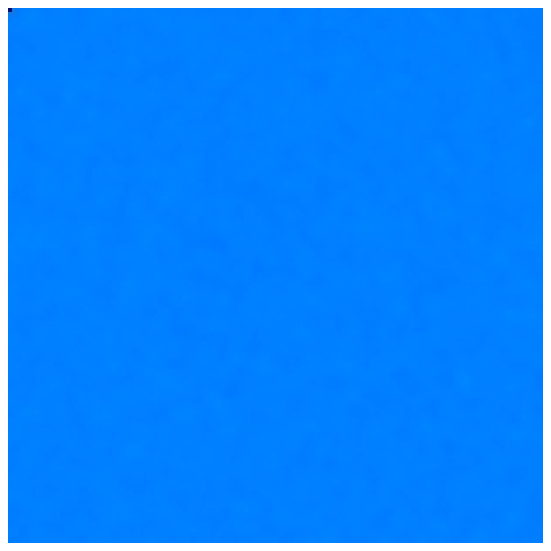}\\
\includegraphics[width=0.1\textwidth]{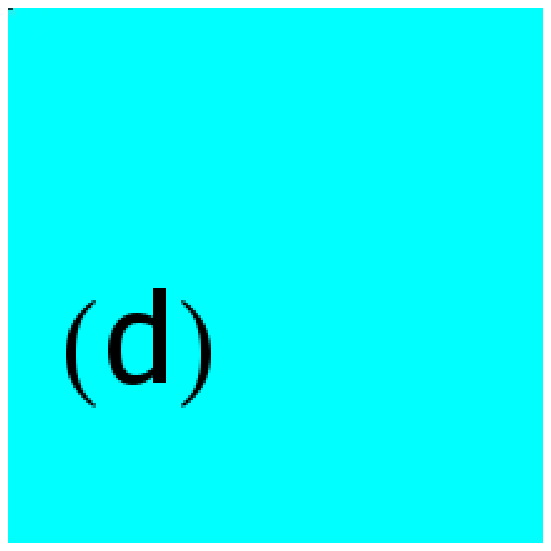}&
\includegraphics[width=0.1\textwidth]{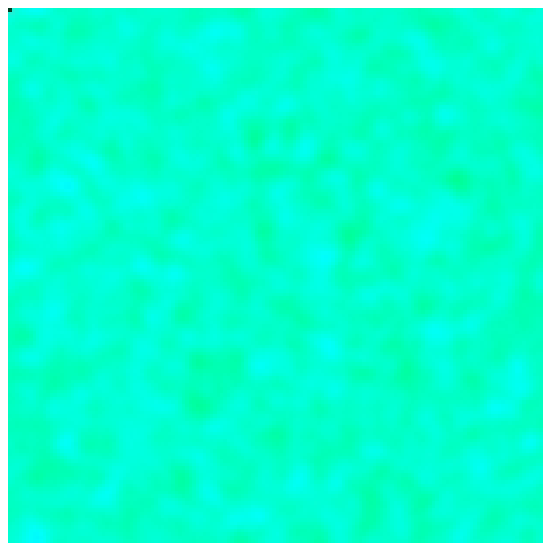} &
\includegraphics[width=0.1\textwidth]{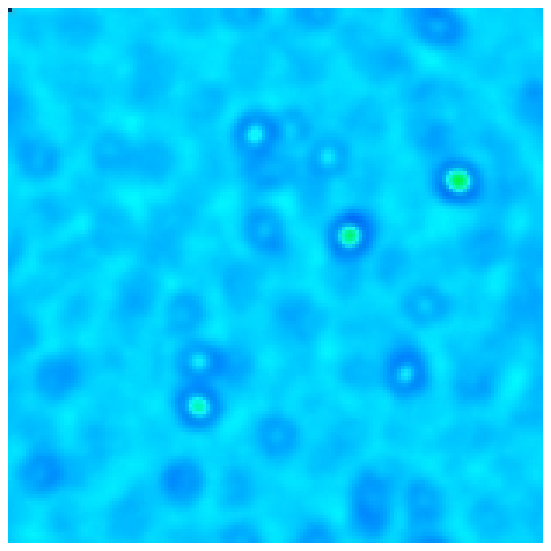} &
\includegraphics[width=0.1\textwidth]{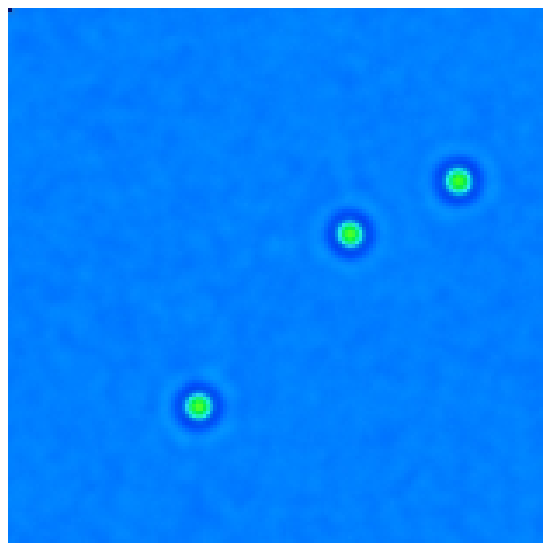}\\
\end{tabular}
\caption{\label{fig}(Color online) Each row of pictures shows -from
left to right- the time evolution  of $\Re{u}(\hat{x},\hat{y},\tau)$
for a different simulation. The initial condition, at $\tau=0$, is
shown on the left most picture, see text for details.  (a) and (b)
have the same initial condition and reaction parameters. (a) If
$\epsilon \rightarrow 0$, which corresponds to the  {\em perfect
mixing regime}, the system evolves to the uniform absorbing state
$R$ (here shown at $\tau=50$). (b) As $\epsilon$ increases, in the
{\em imperfect mixing regime},  the system evolves to a {\em new}
active state with globular replicating structures ($\tau=50$).
Equivalently, for a new initial condition and parameter set, (c)
evolves in the {\em perfect mixing regime}  to the uniform active
state $B_{-}$ ($\tau=300$), whereas (d), in the {\em imperfect
mixing regime}, evolves to a {\em new} active state ($\tau=300$).}
\end{center}
\end{figure}

\begin{figure}[h]
\begin{center}
\includegraphics[width=0.35 \textwidth]{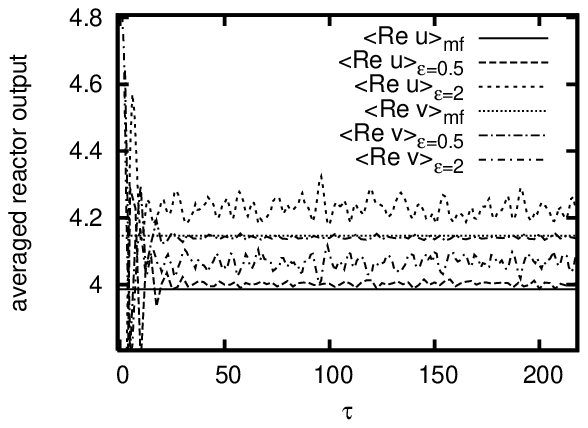}
\includegraphics[width=0.35 \textwidth]{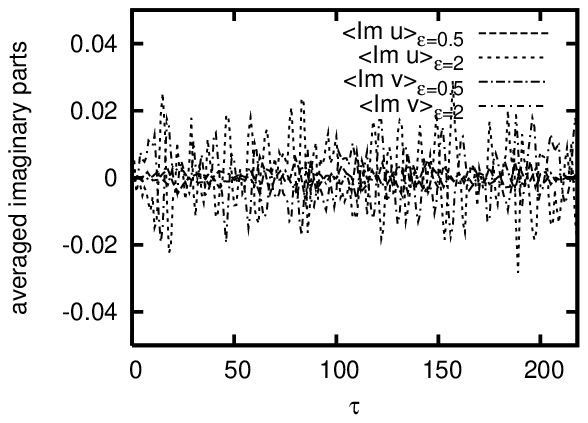}
\caption{\label{fig2}
(Upper) After a transient time, in the perfect mixing limit
$\epsilon=0.5$, the average of the real part of the field tends to
the mean-field value . However, in the imperfect mixing regime,
$\epsilon=2$, these averages deviate significantly from the
mean-field solution and show oscillatory behavior. (Lower) The
averaged imaginary parts of the fields are always negligible since
the averages correspond to physical, real, densities. }
\end{center}
\end{figure}

In Fig. \ref{fig} we display some ``snapshots" of the time evolution
of the real part of the nutrient field $u(\hat{x},\hat{y},t)$ for
two different parameter sets and two diffusion rates. When displayed
in color, blue represents a concentration between $\frac{0.2}{\mu}$
and $\frac{0.4}{\mu}$, green close to $\frac{0.5}{\mu}$, yellow
represents an intermediate concentration of roughly
$\frac{0.8}{\mu}$ and red close to $\frac{1}{\mu}$. On a grayscale,
lighter grays correspond to low concentration and darker ones to
high concentration. Fig.\ref{fig}-(a) and (b) represent the time
evolution of the same initial homogeneous condition
$(u=\frac{1}{\mu},v=\frac{0.1}{\mu})$ for $\mu=0.095$ and
$\nu=0.03$. In case (a),  a regime of high diffusion rates
($\epsilon=0.01225$), initially there are some local spatial
fluctuations about the mean value. Then, the fluctuations are
homogenized due to the high diffusivity of the reactants and the
system evolves towards the inactive homogeneous state  $R$ of the
mean-field approach. This is the result we may expect in a perfect
mixing regime. In case (b), we have increased the noise strength
($\epsilon=0.0225$) by reducing the diffusion rates, i.e. to explore
deviations from the mean-field results we allow for the imperfect
mixing effects: the strong spatial incoherent fluctuations give rise
to spatial patterns with self-replicating and moving globules. In
the interior of each of these units, in blue, there is a region with
sustained autocatalytic production of $v$ which is causing the local
depletion of the substrate $u$.  A similar experiment is shown in
Fig. \ref{fig}-(c) and (d), with a new initial condition
$(u=\frac{0.5}{\mu},v=\frac{0.25}{\mu})$, and parameter set
$\mu=0.11025$, $\nu=0.05$. The initial condition is again the same
in both cases, but in the perfect mixing regime where  the system
has a high diffusion rate (and a low noise intensity such as
$\epsilon=0.01$) it evolves to the uniform stable active state
$B_{-}$ (see case (c)), whereas in the imperfect mixing regime with
low diffusion rates (and higher noise intensities such as
$\epsilon=0.023$) the system evolves to a new active pattern with
globular structures. Thus, in cases (b) and (d)  a low diffusion
rate has induced fluctuations that drive the system away from the
absorbing state $R$ or the uniform blue state $B_{-}$ and produces
spatial compartmentalization.
Due to the imperfect mixing effects the averaged output of the
reactor (integrated over $\hat{x}$ and $\hat{y}$) may also change
with the diffusion rate. In Fig. \ref{fig2}-(upper) we show, for
$\mu=0.0605$, $\nu=0.02$ and starting with the initial homogeneous
condition $(u=\frac{0.3}{\mu},v=\frac{0.25}{\mu})$, how for the
average of the real part of the field, which corresponds to the
density, different mixing regimes lead to different reactor outputs.
For very high diffusion rates (and low noise intensities
$\epsilon$), the system evolves to the mean-field solution
$(\langle\Re u \rangle_{mf},\langle \Re v \rangle _{mf})=B_{-}$,
notice that $\langle \Re v \rangle > \langle \Re u \rangle  $. If
the diffusivity of reactants is reduced the system evolves to a new
equilibrium state with the reversed ratio $\langle \Re v \rangle <
\langle \Re u \rangle$. Furthermore, after the initial transient
time, the  output of the reactor shows {\em oscillatory} behavior
which increases in amplitude as we deviate from the perfect mixing
regime. In Fig.  \ref{fig2}-(lower) we show  for the same
simulations the average of the imaginary part of the field,
confirming that the averaged imaginary part is zero up to the
statistical error.
Thus in any system governed by a nonlinear rate law, a knowledge of
the bulk or average concentrations is not sufficient to predict
neither the average rate of the reaction, nor the final equilibrium
spatial distribution or the averaged reactor output. One must also
know the spatial distribution of the reacting material. Furthermore,
there is a strong dependence on the ratio of the reaction to
diffusion rate, and the mean-field approach will only be valid in
the limit where this ratio vanishes. The noise parameter $\epsilon$
is a function of the spatial dimensionality suggesting that the
resulting rates of reaction, spatial distribution and global
averaged densities of reactants may change from two to three
dimensions and the simple extrapolation of two-dimensional models
results may not be adequate. Finally, these results show that one
can thus tune the mixing efficiently to control the composition  of
the output of the reactor and second, that for certain ranges of
diffusion rates we may discover unexplored dynamical ranges where
spatial organization takes place. One may also appreciate the
relevance that the {\em mixing} process will have in biological
systems, such as epidemics propagation or in viral dynamics
\cite{BE2}, where spatial segregation may facilitate coexistence.

This general-purpose, consistent mathematical and numerical approach
allows us to explore rigorously the effects of reaction induced
fluctuations in imperfect mixing regimes with complex fields and
noises. Dealing with the full system we observe that there is
multiplicative noise acting on {\em both} fields. If we had ignored
the zero-autocorrelated complex field $\eta$ in
Eq.(\ref{noisecorr}), then only the $v$ field would have had noise
and both field and $\xi$ would have been real \cite{CPL}. This
approximation is not correct in the general case and the system
outcome may be significantly different in non-linear systems such as
this one, which is quite sensitive to small fluctuations.

M.-P. Z. is supported by the Instituto Nacional de T\'{e}cnica
Aerospacial (INTA). The research of D.H. is supported in
part by INTA and F.M. is supported in part by grant
BMC2003-06957 from MEC (Spain).

\end{document}